# All-electric mimicking synaptic plasticity based on the noncollinear antiferromagnetic device

*Cuimei Cao[1], Wei Duan[1], Xiaoyu Feng[2], Yan Xu[1], Yihan Wang[1], Zhenzhong Yang[2], Qingfeng Zhan[2], and Long You[1, 3, 4\*]*

1 School of Integrated Circuits & Wuhan National Laboratory for Optoelectronics, Huazhong University of Science and Technology, Wuhan 430074, China

2 Key Laboratory of Polar Materials and Devices (MOE), School of Physics and Electronic Science, East China Normal University, Shanghai 200241, China

3 Key Laboratory of Information Storage System, Engineering Research Center of data storage systems and Technology, Ministry of Education of China, Huazhong University of Science and Technology, Wuhan, 430074, China

4 Shenzhen Huazhong University of Science and Technology Research Institute, Shenzhen, 518000, China

\* Authors to whom correspondence should be addressed: lyou@hust.edu.cn

**ABSTRACT**

**Neuromorphic computing, which seeks to replicate the brain's ability to process information, has garnered significant attention due to its potential to achieve brain-like computing efficiency and human cognitive intelligence. Spin-orbit torque (SOT) devices can be used to simulate artificial synapses with non-volatile, high-speed processing and endurance characteristics. Nevertheless, achieving energy-efficient all-electric synaptic plasticity emulation using SOT devices remains a challenge. We chose the noncollinear antiferromagnetic $Mn_3Pt$ as spin source to fabricate the $Mn_3Pt$-based SOT device, leveraging its unconventional spin current resulting from magnetic space breaking. By adjusting the amplitude, duration, and number of pulsed currents, the $Mn_3Pt$-based SOT device achieves nonvolatile multi-state modulated by all-electric SOT switching, enabling**

**emulate synaptic behaviors like excitatory postsynaptic potential (EPSP), inhibitory postsynaptic potential (IPSP), long-term depression (LTD) and the long-term potentiation (LTP) process. In addition, we show the successful training of an artificial neural network based on such SOT device in recognizing handwritten digits with a high recognition accuracy of 94.95 %, which is only slightly lower than that from simulations (98.04 %). These findings suggest that the $Mn_3Pt$-based SOT device is a promising candidate for the implementation of memristor-based brain-inspired computing systems.**

Neuromorphic computing is an emerging computing paradigm that mimics the operation of a biological brain at different levels of computational hierarchy.[1-4] Unlike traditional computing systems following the von Neumann architecture, which involve continuous data exchange between memory units and the central processing unit, neuromorphic computing processes information in a decentralized manner. This approach reduces energy consumption and enhances efficiency in handling complex cognitive tasks such as recognition, reasoning, and interaction. Despite significant progress in artificial neural network algorithms, the development of neuromorphic computing has been hindered by the lack of specialized hardware. As a result, various materials and structures, including phase change materials,[5, 6] ferroelectric materials,[7, 8] and resistive random-access memory[9, 10] and spintronic device have been proposed to address this challenge.

Spintronic devices are distinguished from other competitors due to their non-volatile, low energy consumption, high-speed processing and endurance characteristics.[11-15] Additionally, spintronic devices exhibit essential traits required for neuromorphic computing, such as nonlinearity, stochasticity, and nonvolatility.[4, 13, 14] Present works in spintronic neuromorphic computing focus on electrically manipulating the magnetization of a ferromagnet using spin-orbit torque (SOT).[15, 16] In a typical SOT device, an in-plane charge current ($I$) flowing along the $x$-direction in a heavy metal (HM) layer generates spin currents with orthogonal spin polarization ($\sigma_y$, $y$-spins) through spin-orbit coupling. The spin currents flow in the $z$-direction and apply an in-plane SOT on the magnetization of an adjacent ferromagnetic layer to control it.[17-19] However, in-plane SOT is ineffective and random for switching the magnetization with perpendicular magnetic anisotropy (PMA). Thus, an external assisted magnetic field ($H_x$) along the current direction is necessary to break the mirror symmetry of the perpendicular magnetization. To avoid the necessity of such magnetic field, various approaches (such as exchange bias fields, tilted magnetization anisotropy, asymmetric structures, interlayer exchange coupling, composition gradients, external electric fields, and so on)[20-28] have been implemented to disrupt the mirror symmetry of the perpendicular magnetization between up and down magnetization states, and hence achieve SOT field-free switching, namely all-electric SOT switching. Among these techniques, the spin current with out-of-plane spin polarization ($\sigma_z$, $z$-spins) can directly generate out-of-plane SOT, overcoming the symmetry constraints of

SOT devices with PMA and enhancing the efficiency of SOT-induced magnetization switching.[29-32] This method proves to be an effective means of controlling perpendicular magnetization and has garnered increasing attention. However, the SOT efficiency of material generating $z$-polarized spin current is generally smaller than that of HM. Recently, many works reported that the noncollinear antiferromagnet (such as $Mn_3GaN$,[33] $Mn_3Sn$,[32] $Mn_3Pt$[34, 35] and $Mn_3Ir$[36, 37]) can be identified as an exceptional spin source capable of producing $z$-polarized spin current while demonstrating high SOT efficiency.

$Mn_3Pt$,[38, 39] a noncollinear antiferromagnet with cubic structure and Mn atom align on kagome (111) plane, has drawn considerable attention due to its unique physical properties, such as anomalous Hall effect,[34, 40] spin Hall effect,[34] magnetic spin Hall effect,[34] out-of-plane spin polarization,[35] antiferromagnetic tunneling magnetoresistance effect,[41] etc. In our previous work, we have realized that the all-electrical control of magnetization states of "up" and "down" by out-of-plane spin torque from $Mn_3Pt$, which corresponds to "0" and "1" states in the SOT magnetic random-access memory (MRAM).[34] However, the nonvolatile multi-state controlled by all-electric SOT switching in $Mn_3Pt$-based SOT device is rarely reported, which is a crucial issue for the neuromorphic computing applications.

In this work, we present all-electric SOT switching with nonvolatile multi-state behaviors in a $Mn_3Pt$/Ti/CoFeB/MgO heterostructure. The device's synaptic weight, represented by the anomalous Hall resistance ($R_H$), can be finely adjusted in an analog manner through manipulating of the pulsed amplitude and duration. This behavior mirrors the stimulus-dependent modulation observed in biological synapses. Furthermore, the $Mn_3Pt$-based SOT-device exhibits several functionalities akin to biological synapses, such as excitatory postsynaptic potential (EPSP), inhibitory postsynaptic potential (IPSP), long-term depression (LTD) and the long-term potentiation (LTP) processes for the advancement of neuromorphic computing. Employing $Mn_3Pt$/Ti/CoFeB/MgO heterostructure as a synapse, an artificial neural network (ANN) established for handwritten digit recognition is further simulated, exhibiting a high recognition rate (94.95 %) comparable to simulation result (98.04 %). This work identifies noncollinear antiferromagnet $Mn_3Pt$-based SOT device as an intriguing candidate for implementing the neuromorphic spintronics with high energy efficiency.

**RESULTS AND DISCUSSIONS**

**Crystal Characterization of Mn$_3$Pt Films**

The (001)-oriented Mn$_3$Pt film was deposited on MgO(001) substrate using a magnetron sputtering system and the growth detail can be found in our previous work.[34] As shown in Figure 1a, the noncollinear antiferromagnetic material Mn$_3$Pt, which possesses a cubic structure, belongs to the space group $Pm\bar{3}m$. In this structure, the Mn moments are arranged in either "head-to-head" or "tail-to-tail" noncollinear alignments within the (111) kagome planes. Here, we have successfully grown high-quality epitaxial Mn$_3$Pt films on single-crystalline MgO(001) substrates, which can be verified through X-ray diffraction (XRD) with a Bruker D8 Discover diffractometer using Cu $K\alpha$ radiation ($\lambda$ = 0.15419 nm) and cross-sectional aberration-corrected scanning transmission electron microscope (AC-STEM, FEI Titan Themis 200) techniques. Figure 1b shows the out-of-plane XRD $\theta$-$2\theta$ pattern of the MgO(001)//Mn$_3$Pt(15 nm) sample, indicating a high degree of (001) texture emerges in the Mn$_3$Pt film. As illustrated in Figure 1c, $\phi$ scanning around the (111) peaks of the Mn$_3$Pt film and MgO substrate confirms the four-fold symmetry of the Mn$_3$Pt film well matching to that of the MgO substrate. In addition, AC-STEM further revealing high-quality epitaxy with a sharp boundary between the Mn$_3$Pt film and the MgO substrate, as shown in Figure 1d.

**Magnetization Switching in Mn$_3$Pt-based SOT device**

To investigate the SOT-induced magnetization switching and memristor-like behavior in Mn$_3$Pt, we grew the multilayers consisting of Mn$_3$Pt(8)/Ti(3)/CoFeB(1)/MgO(2)/SiO$_2$(3) on MgO substrate (the numbers in brackets are the nominal thicknesses in nanometers, from bottom side) using DC/RF magnetron sputtering technique and more growth detail can be found in our previous work.[34] For electrical transport measurements, the multilayers were patterned into a Hall bar structure by conventional photolithography and Ar-ion-beam etching. Then, the Ti(5 nm)/Cu(100 nm) as four top electrodes of Hall bar was deposited by DC magnetron sputtering. The schematic representation of the measurement setup and stack structure are illustrated in Figure 2a. The anomalous Hall effect and the SOT-induced magnetization switching of Mn$_3$Pt/Ti/CoFeB/MgO heterostructures were measured using a homemade instrument using Keithley 6221 (current source) and 2182 (nanovoltmeter) combination. The anomalous Hall resistances ($R_H$) were recorded while sweeping the out-of-plane magnetic field ($H_z$) at room temperature. The squareness of the anomalous Hall effect (AHE) loop indicates the excellent PMA of the CoFeB layer, as shown in Figure 2b, which enables us to monitor the magnetization state of the CoFeB layer by the $R_H$ during the SOT-induced magnetization switching. We measured the SOT-induced magnetization switching under different in-plane magnetic fields ($H_x$), as shown in Figure 2c. The switching polarity was reversed when $H_x$ with opposite direction was applied, indicating the current control of magnetization was governed by SOT.

A remarkable and deterministic magnetization switching with a 200 μs pulse current was achieved at $H_x$ = 0 Oe, that is all-electric SOT switching, as shown in Figure 2c. The phenomenon of all-electric SOT switching can be attributed to the z-polarized spin current in Mn$_3$Pt, which results from its low-symmetry magnetic group.[34] This represents an efficient method for all-electrically manipulating the perpendicular magnetization. Moreover, it has been observed that the corresponding switching ratio of the CoFeB layer utilizing a Mn$_3$Pt spin source is approximately 80 %, surpassing that of a Co/Ni multilayer driven by Mn$_3$Sn layer (≈ 60 %).[32] Here, the SOT-induced magnetization switching ratio is defined as $\Delta R_H/R_H$, where $\Delta R_H$ and $R_H$ represent the current-induced and field-induced change of the anomalous Hall resistance respectively. Additionally, the critical switching current density for all-electric SOT switching of CoFeB in Mn$_3$Pt/CoFeB/MgO heterostructures ($J_{Mn_3Pt}$) is 9.40×10$^6$ A/cm$^2$, which is smaller than that of the heavy metal with field-assisted magnetization switching (for Pt/CoFeB/MgO,[42] $J_c$ = 2×10$^7$ A/cm$^2$; for Ta/CoFeB/MgO,[43] $J_c$ = 1.8×10$^7$ A/cm$^2$), collinear antiferromagnet[23] (the $J_c$ for Ta/IrMn/CoFeB/MgO and CoFeB/IrMn/CoFeB are 2.67×10$^7$ A/cm$^2$ and 3.09×10$^7$ A/cm$^2$, respectively), and other noncollinear antiferromagnet (for Mn$_3$Ir,[37] $J_c$ = 1.48×10$^7$ A/cm$^2$), suggesting that the Mn$_3$Pt represents an ideal spin source material for energy-efficient magnetization switching of ferromagnet. Note that the current density $J_{Mn_3Pt}$ in the Mn$_3$Pt layer has been evaluated by taking into account the shunting effect of the Ti and CoFeB layers in the device (Note 1 in the Supporting Information).

**Mimicking Synaptic Behaviors**

To further explore the potential of Mn$_3$Pt/Ti/CoFeB/MgO heterostructure as efficient artificial synapses, the variation of AHE resistance $R_H$ during the all-electric SOT switching were further characterized by adjusting the amplitude/duration of current pulses. Figure 3a illustrates all-electric SOT switching in the Mn$_3$Pt/Ti/CoFeB/MgO heterostructure with varied pulse widths ranging from 50 to 400 μs. Both the critical switching current density $J_{Mn_3Pt}$ and the number of distinguishable intermediate states exhibits increase, as the pulse width decreases. This type of memristor-like behavior is essential for its application in neuromorphic computing. Figure 3b displays $\Delta I$ as a function of pulse width required for all-electric SOT switching, where $\Delta I$ denotes the minimum current range for the magnetization switching process, as depicted in the inset of Figure 3b. It is evident that $\Delta I$ demonstrates a non-linear increase as the pulse width decreases, suggesting that the number of intermediate states can be precisely controlled by the pulse width, thereby exhibiting significant plasticity. Subsequently, if one varies the maximal magnitude of the current density, $J_{max}$, from low (8.0×10$^6$ A/cm$^2$) to high (11.5×10$^6$ A/cm$^2$), a series of hysteretic loops are observed, as illustrated in Figure 3c, indicating that the final resistance state can effectively be determined by the current cycle protocol. The presented curves distinctly illustrate the presence of multiple non-volatile AHE resistances as the current density is progressively increased. This impulse-regulated multi-state response in Mn$_3$Pt device, which is a typical memristor-like behavior, will be used for mimicking the behaviors of artificial synapses, such as excitatory postsynaptic potential (EPSP) and inhibitory postsynaptic potential (IPSP) processes.

Then, the characteristics of resistance variations were employed through programming of consecutive pulse sequences to investigate the synaptic behavior of the Mn$_3$Pt-based SOT device. For this measurement, the magnetization of CoFeB layer firstly was set into the up state by a positive reset current pulse ($J_{Mn_3Pt}$= 15×10$^6$ A/cm$^2$), and then a series of negative and positive current pulses with pulse number/width of 50/200 μs were applied. The fixed current pulse amplitude is 8.5×10$^6$ A/cm$^2$, which corresponds to the very beginning switching points of the device. As illustrated in Figure 3d, when the number of positive/negative current pulses increases, the $R_H$ increases (marked by purple points)/decreases (marked by red points) gradually. The response of $R_H$ to these consecutive pulse sequences (Figure 3d) indicates that either positive or negative pulses with the same properties (such as amplitude, duration, and number) can reproducibly determine the almost same resistance state (~0.3 Ω), corresponding to the synaptic plasticity of IPSP or EPSP, respectively.

Additionally, LTD and LTP are another key characteristic of synapses. Through applying pulses with suitable pulse amplitudes, we could regulate the linear LTP and LTD processes of the device's synaptic weight. During this measurement, a positive/negative pulse current density ranging from $8.0\times10^6$ A/cm$^2$ to $12.0\times10^6$ A/cm$^2$ with a step of $0.03\times10^6$ A/cm$^2$ is applied to the device for investigating synaptic plasticity. As shown in Figure 4a, the Hall resistance $R_H$ demonstrates a nearly linear increase or decrease in response to the application of 150 positive or negative pulses, respectively. More importantly, it displays symmetrical behavior in both polarities, effectively reflecting the synaptic activities associated with LTP and LTD. Furthermore, a performance benchmark of such magnetization switching in comparison with those investigated in other SOT synaptic devices with all-electric SOT switching characteristics was conducted (Table 1). Compared to other devices, the Mn$_3$Pt/Ti/CoFeB/MgO heterostructure exhibited a notably high switching ratio (approximately 80 %) while maintaining low critical switching current density ($9.40\times10^6$ A/cm$^2$), which enhances its attractiveness for potential neuromorphic applications.

**ANN System with Mn$_3$Pt-based artificial synapse**

Exploiting the measured characteristics of artificial synapse using Mn$_3$Pt-based SOT device (Figure 4a), we developed a synapse model in Python and constructed a three-layer fully connected ANN with a 784-100-10 architecture for handwritten digit recognition, as depicted in Figure 4b. The neurons in the neural network employ the ReLU activation function, while three distinct types synapses were adopted for comparison: software-based synapses, defect-free synaptic devices, and synaptic devices with non-ideal characteristics. The non-ideal characteristics include device-to-device variations, cycle-to-cycle variations, and readout variations among devices. We carried out the write and read endurance test by applying pulse currents with 200 μs duration to the Mn$_3$Pt-SOT device (Note 2 in Supporting Information). As illustrated in Figure S1a, the write endurance measurement shows that the low/high AHE resistance obtained by alternately applying positive/negative pulses remained nearly unchanged after 180 repetitions, indicating the excellent stability of Mn$_3$Pt-based SOT device. In addition, for read endurance test, a pulse current of 15×10$^6$ A/cm² is initially applied into the Hall bar device and subsequently removed. The $R_H$ is continuously read for 200 times, demonstrating excellent reading endurance of the Mn$_3$Pt/Ti/CoFeB/MgO heterostructure (Figure S1b in Supporting Information). Furthermore, the non-ideal characteristics in the Python model are derived from results obtained by measuring multiple SOT devices (Note 3 in Supporting Information). As shown in Figure S2, the experimentally measured LTP and LTD processes of 10 Mn$_3$Pt-based synaptic devices all exhibit linearity. Combining the endurance test results of writing and reading, it is demonstrated that the Mn$_3$Pt/Ti/CoFeB/MgO heterostructure is an intriguing candidate for neuromorphic spintronics. Figure 4c presents the in-situ training results of the ANNs utilizing the three distinct types of synapses, where the network employing ideal software-based synapses achieves the highest recognition accuracy of 98.04 %. Importantly, the accuracy of the networks with non-ideal devices decreased only slightly to 94.95 %, while it is around 2 % higher than that of the network with defect-free devices (92.95 %). These results imply that for the Mn$_3$Pt-based SOT synapse, training with errors can indeed enhance the robustness and error tolerance of the artificial neural network. Moreover, in Figure 4d, we give the evolution of the Mn$_3$Pt-based SOT synapse weight matrices of the input-hidden layer (W1) and hidden-output layer (W2) before and after training. Evidently, the weights of the entire network have been significantly modified with training of the SOT synapses, revealing the weights distribution becomes more regular and directional after training. These experimental results demonstrate that the SOT device based on noncollinear antiferromagnet Mn$_3$Pt is considered a promising candidate for exploring the potential applications of memristor-like devices in neuromorphic computing.

**CONCLUSIONS**

In summary, the nonvolatile multi-state of CoFeB ferromagnet were modulated by unconventional spin torque from a noncollinear antiferromagnetic Mn$_3$Pt without any external assisted magnetic field. The realization of low critical current density and notably large all-electric SOT switching ratio in Mn$_3$Pt/Ti/CoFeB/MgO heterostructures prove the superior efficiency of the Mn$_3$Pt-based SOT device for emulating the synaptic behavior used in the neuromorphic computing. The synaptic weight, represented by $R_\mathrm{H}$, can be effectively manipulated by adjusting the amplitude and number of pulsed current. Moreover, the memristive behavior in the Mn$_3$Pt-based SOT device can be utilized to simulate synaptic behaviors, such as EPSP, IPSP, LTD and LTP. By employing the Mn$_3$Pt-based SOT device as a synapse, an ANN is constructed to recognize handwritten digits with a high recognition rate (94.95 %), which is comparable to soft-based network (98.04 %). This demonstration highlights the Mn$_3$Pt-based SOT heterostructure had good potential for emulating neuromorphic devices to construct energy-efficient neural networks.

## ASSOCIATED CONTENT

**Supporting Information**

The Supporting Information is available free of charge.

Determination of the current density of $Mn_3Pt$ layer; Write and read endurance test in $Mn_3Pt$/Ti/CoFeB/MgO system; The measurement of LTD and LTP processes of synaptic weight for 10 devices.

## AUTHOR INFORMATION

**Corresponding Author**

**Long You** − *School of Integrated Circuits & Wuhan National Laboratory for Optoelectronics, Huazhong University of Science and Technology, Wuhan 430074, China; Key Laboratory of Information Storage System, Engineering Research Center of data storage systems and Technology, Ministry of Education of China, Huazhong University of Science and Technology, Wuhan, 430074, China; Shenzhen Huazhong University of Science and Technology Research Institute, Shenzhen, 518000, China*; Email: lyou@hust.edu.cn

**Authors**

**Cuimei Cao** − *School of Integrated Circuits & Wuhan National Laboratory for Optoelectronics, Huazhong University of Science and Technology, Wuhan 430074, China*

**Wei Duan** − *School of Integrated Circuits & Wuhan National Laboratory for Optoelectronics, Huazhong University of Science and Technology, Wuhan 430074, China*

**Xiaoyu Feng** − *Key Laboratory of Polar Materials and Devices (MOE), School of Physics and Electronic Science, East China Normal University, Shanghai 200241, People's Republic of China*

**Yan Xu** − *School of Integrated Circuits & Wuhan National Laboratory for Optoelectronics, Huazhong University of Science and Technology, Wuhan 430074, China*

**Yihan Wang** − *School of Integrated Circuits & Wuhan National Laboratory for Optoelectronics, Huazhong University of Science and Technology, Wuhan 430074, China*

**Zhenzhong Yang** – *Key Laboratory of Polar Materials and Devices (MOE), School of Physics and Electronic Science, East China Normal University, Shanghai 200241, People's Republic of*

*China;*

**Qingfeng Zhan** − *Key Laboratory of Polar Materials and Devices (MOE), School of Physics and Electronic Science, East China Normal University, Shanghai 200241, People's Republic of China;*

**Author Contributions**

C.M.C. and L.Y. conceived the idea and supervised the project. C.M.C. grew sample, fabricated the Hall bar device and performed the XRD measurements. C.M.C., Y.X. and Y.H.W. performed electrical transport measurements. X.Y.F and Z.Z.Y. preformed the AC-STEM measurement. W.D. performed the simulations. C.M.C. analyzed the results together with Q.F.Z. and L.Y.. All authors contributed to discussions. C.M.C. and L.Y. wrote the manuscript.

**Notes**

The authors declare no competing financial interest.

**ACKNOWLEDGEMENTS**

This work supported by the National Natural Science Foundation of China (NSFC Grant Nos. 52401301, 12327806, 62074063, 61821003), the China Postdoctoral Science Foundation (No. 2024M6750982), Shenzhen Science and Technology Program (Grant No. JCYJ20220818103410022), National Key Research and Development Program of China (Grant No. 2020AAA0109005, 2023YFB4502100).

**Table 1**. A performance benchmark of $Mn_3Pt$-based SOT switching in comparison with those investigated in other SOT-devices

| SOT-based neuromorphic device | All-electric SOT switching through | Switching ratio $\Delta R_H/R_H$ | Critical current density $J_c$ ($\times 10^6$ A/cm$^2$) | Ref. |
|---|---|---|---|---|
| $Mn_3Pt$/Ti/CFB/MgO | Low spin order symmetry | 80 % | 9.4 | This work |
| CuPt/CoPt | Low crystal symmetry | 69.1 % | 50 | 44 |
| FePt/Ti/NiFe | Interlayer exchange coupling | 6~7 % | 5.1 | 45 |
| IrMn/Co/Ru/CoPt/CoO | Interlayer exchange coupling | 42 % (200 K) | 29.8 | 46 |
| W/Pt/Co/NiO/Pt | Exchange bias | 8~10 % | 27~34 | 47 |
| [CoTb]/Pt/$Si_3N_4$ | Composition gradient | 0.78 % | 50~60 | 48 |
| Pt/Co/IrMn | Exchange bias | 25 % | 3 | 49 |
| Pt/Co/Pt(wedged) | Wedged structure | 78.6 % | 64 | 50 |

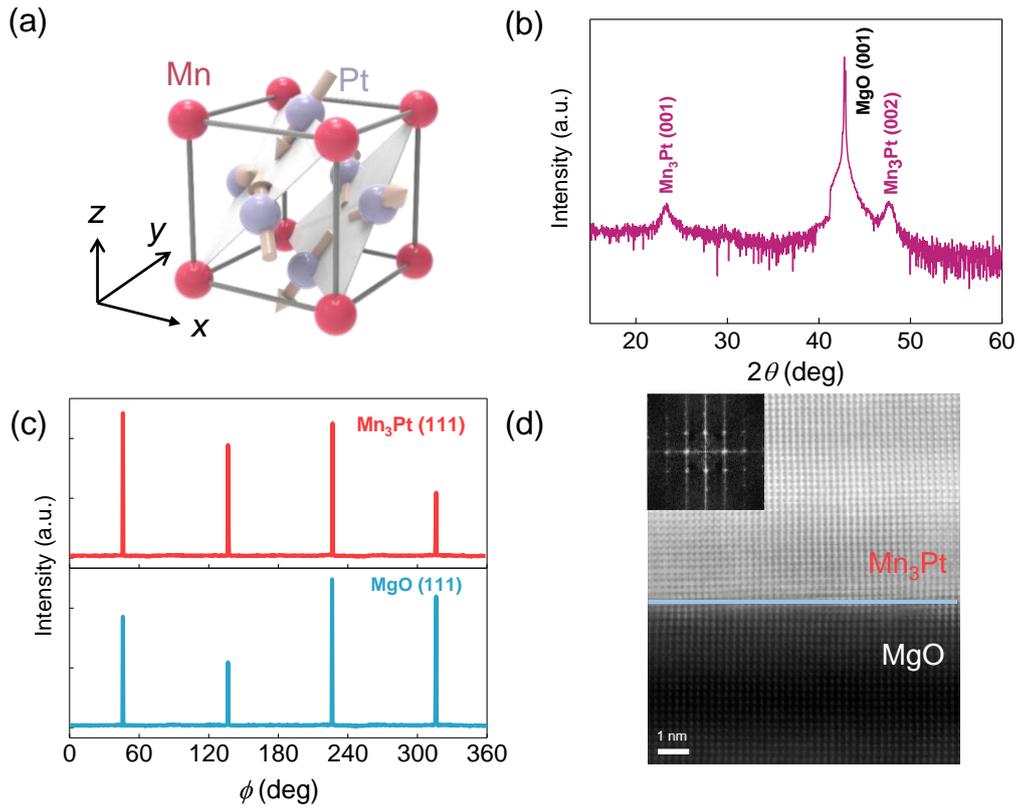

**Figure 1**. Crystal structure characterization of Mn$_3$Pt. (a) The schematic diagram of crystal and magnetic structure of cubic noncollinear antiferromagnet Mn$_3$Pt, where the Mn moments form "head-to-head" or "tail-to-tail" noncollinear alignments in (111) kagome planes, and the nonmagnetic metal Pt is located in the face-centered position of the cubic structure. (b) The representative high-resolution x-ray diffraction $\theta$-$2\theta$ pattern of Mn$_3$Pt film deposited on MgO (001) substrate. (c) $\phi$ scanning patterns of the Mn$_3$Pt film and MgO substrate which taken around the (111) diffraction peaks. (d) The cross-sectional transmission electron microscopy image of an interfacial region of the Mn$_3$Pt/MgO sample, the blue line denotes the sharp interface between Mn$_3$Pt film and MgO substrate.

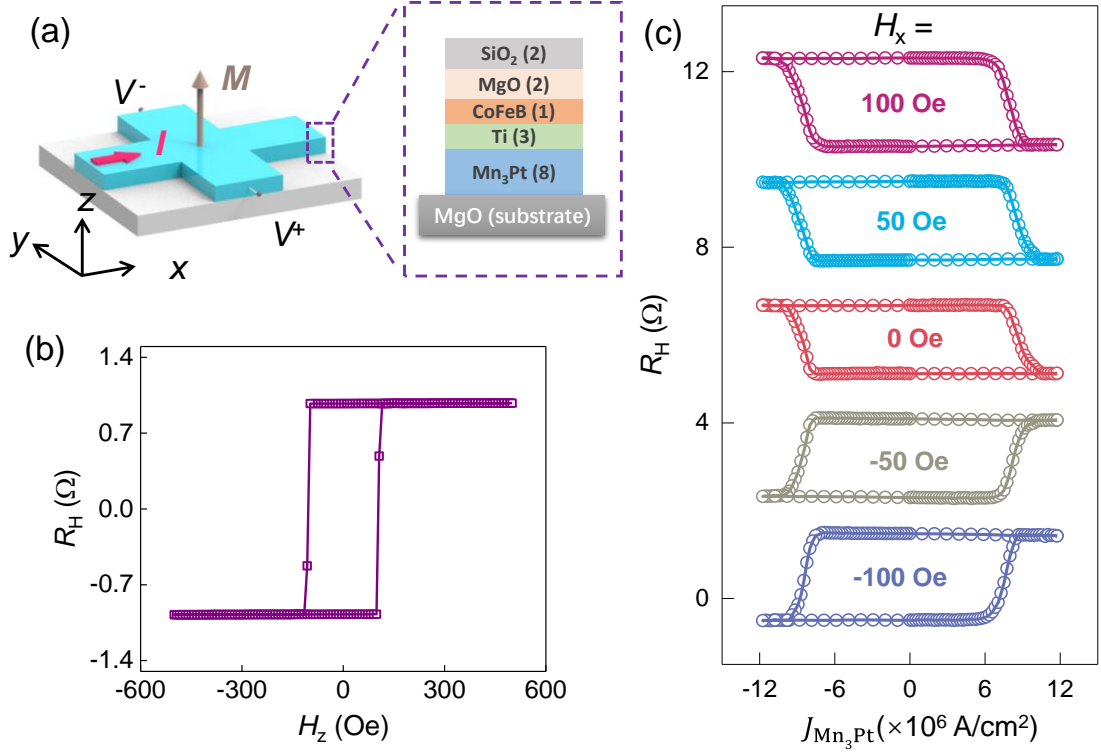

**Figure 2**. All-electric SOT switching in $Mn_3Pt$-based SOT device. (a) The schematic of MgO(001)/$Mn_3Pt$/Ti/CoFeB/MgO/$SiO_2$ heterostructure with the thickness in nanometers and the measurement set-up for anomalous Hall resistance and the SOT-induced magnetization switching, with a current applied along $x$ direction. (b) The anomalous Hall resistance $R_H$ as function of the out-of-plane magnetic field $H_z$ for the MgO(001)/$Mn_3Pt$/Ti/CoFeB/MgO heterostructure. (c) $R_H$ versus current density ($J_{Mn_3Pt}$) under different $H_x$ from 100 to -100 Oe for $Mn_3Pt$/Ti/CoFeB/MgO heterostructure, noting that arrows mean the sweeping direction during the measurements.

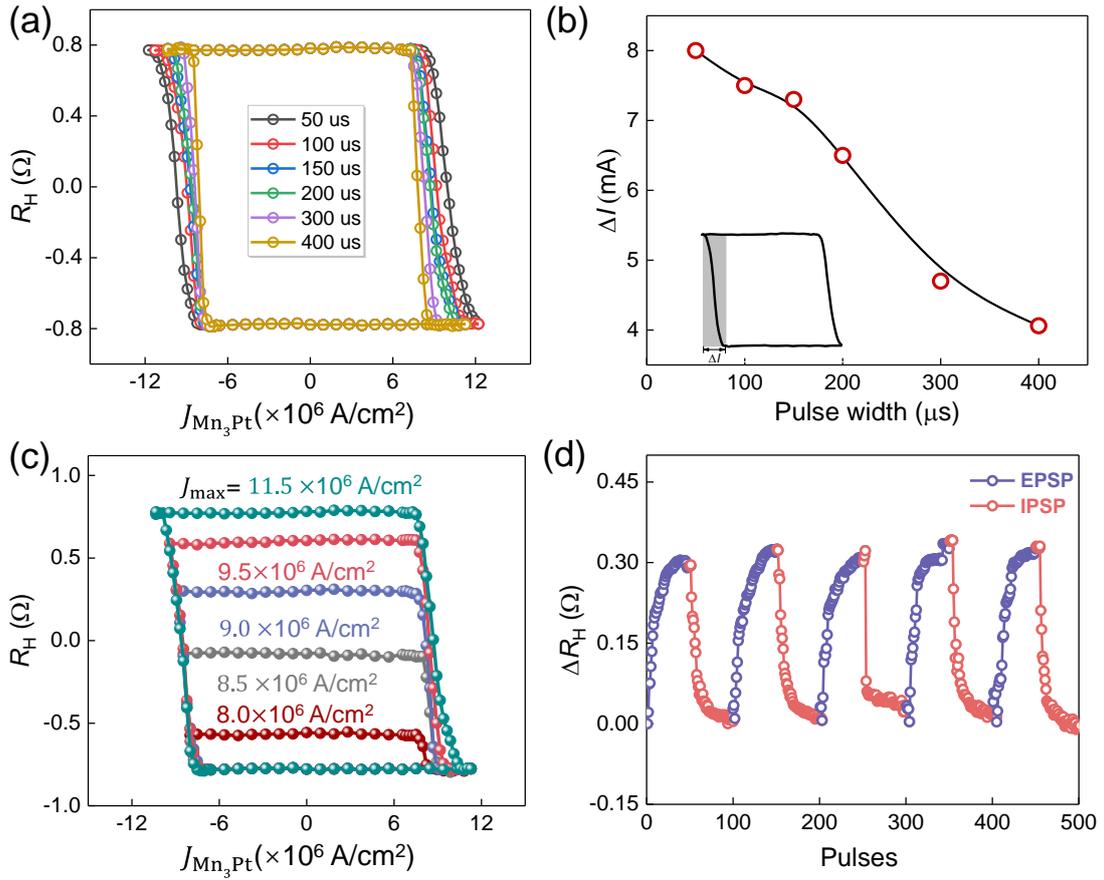

**Figure 3**. Realization of synaptic characteristics by programming consecutive pulses sequences without any assistant magnetic field. (a) All-electric SOT switching loops with different pulses width of current. (b) Tunable current range ($\Delta I$) versus pulse widths of current. (c) All-electric SOT switching measurements were conducted by varying magnitudes of maximal current density ($J_{max}$). (d) The evolution of the excitatory postsynaptic potential (EPSP) and inhibitory postsynaptic potential (IPSP) with consecutive pulses, where 50 positive ($8.5×10^6$ A/cm$^2$) and negative ($-8.5×10^6$ A/cm$^2$) trains pulse currents are alternately applied and the duration is 200 μs.

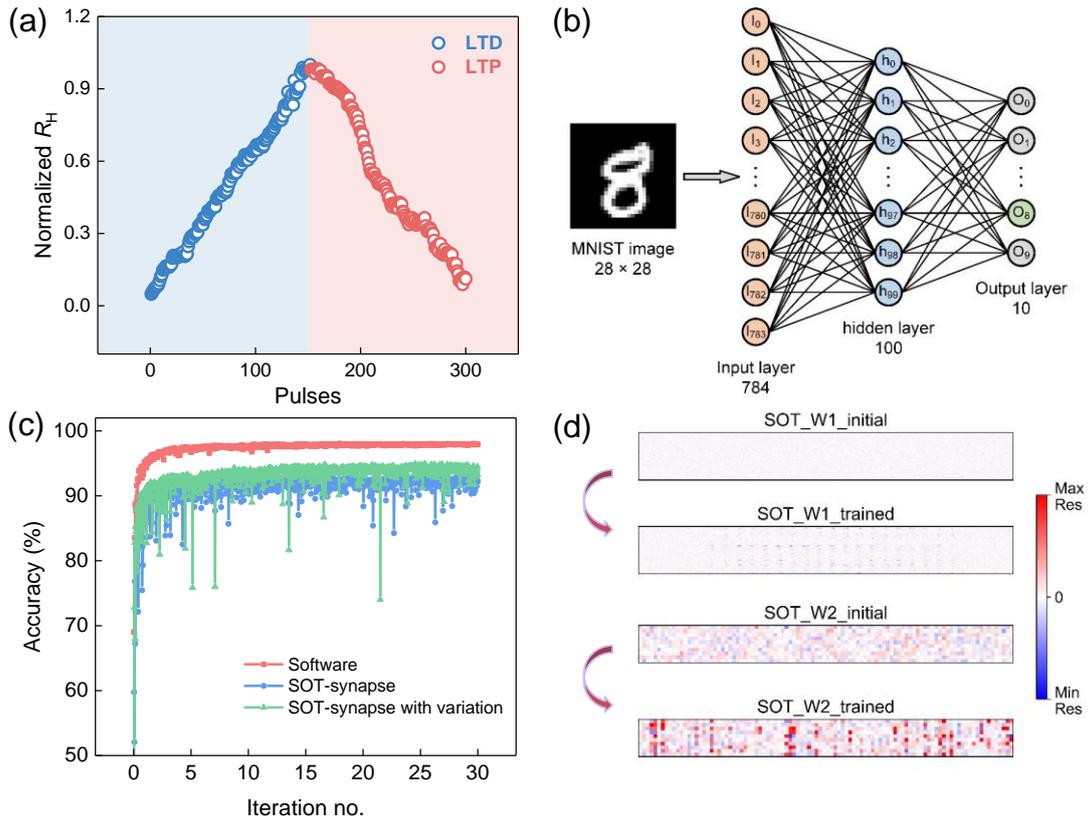

**Figure 4**. The artificial neural network (ANN) established for handwritten digit recognition. (a) long-term depression (LTD) and the long-term potentiation (LTP) process with linearity in $Mn_3Pt/Ti/CoFeB/MgO$ heterostructure. (b) The network structure of the three-layer ANN (784-100-10) for classifying MNIST handwritten digits. (c) Recognition accuracy increases over the course of training. The simulation curve of devices with non-ideal properties (with the accuracy of 94.95 %) follows the software-based (with the accuracy of 98.04 %) one, with a ~3 % gap, and exhibits higher accuracy than that of defect-free devices (92.95 %)). (d) Evolution of the synapse weights of input-hidden layer (W1) and hidden-output layer (W2) before and after in-situ training for the $Mn_3Pt/Ti/CoFeB/MgO$ heterostructure.